\documentclass[preprint,12pt]{elsarticle}
\usepackage{amssymb}
\usepackage{graphicx}

\begin{document}

\begin{frontmatter}

\title{Energetics and stability of vacancies in carbon nanotubes}

\author[usp]{J. E. Padilha}
\ead{padilha@if.usp.br}
\address[usp]{Instituto de F\'{\i}sica, Universidade de S\~ao Paulo, CP 66318, 05315-970, S\~ao Paulo, SP, Brazil}
\author[abc]{R. G. Amorim}
\address[abc]{Centro de Ci\^encias Naturais e Humanas, Universidade Federal do ABC, Santo Andr\'e, SP Brazil}
\author[abc]{A. R. Rocha}
\author[usp]{A. J. R. da Silva}
\author[usp]{A. Fazzio}
\ead{fazzio@if.usp.br}


\date{\today}

\begin{abstract}
In this work we present {\it ab initio} calculations of the formation energies and stability of different types of
multi-vacancies in carbon nanotubes. We demonstrate that, as in the case of graphene, the reconstruction of the defects
has drastic effects on the energetics of the tubes. In particular, the formation of pentagons eliminates the 
dangling bonds thus lowering the formation energy. This competition leads to vacancies having an even number of carbon
atoms removed to be more stable. Finally the appearance of magic numbers indicating more stable defects can 
be represented by a model for the formation energies that is based on the number of dangling bonds of 
the unreconstructed system, the pentagons and the relaxation of the final form of the defect formed after the relaxation.
\end{abstract}


\end{frontmatter}

\section{Introduction}\label{intro}

Nanoscopic systems have attracted significant attention from the scientific community due 
to the possibility of designing ever-smaller electronic devices.  Amongst candidates with 
greatest potential for application one can find carbon-based materials such as carbon 
nanotubes (CNT)\cite{iijima,raman_dresselhaus}, and more recently graphene\cite{Ref01}.

Intrinsic defects are often seen as the source of deleterious effects in semiconductor 
materials\cite{sokrates}. For instance, vacancies and clusters of vacancies have important 
well known effects on the properties of many semiconductors of technological importance such 
as $Si$\cite{Ref18}, $GaAs$\cite{Ref19}, $SiGe$\cite{si_ge_fazzio1,si_ge_fazzio2} and $Ge$\cite{ge_fazzio}. 
The same, however, cannot be clearly stated about carbon based materials. In fact, defects in 
carbon nanotubes can be used as binding sites for different types of gaseous species in CNT-based 
sensors\cite{terrones,rochaprl}, as well as in a new family of disordered graphene spintronics 
devices\cite{rochaprl09}. One point is clear: defects can lead to drastic changes in the electronic 
structure of carbon-based systems. Thus if these materials are to be the building blocks of 
tomorrow's electronics, characterizing these defects is of utmost importance.

In carbon nanotubes one can find many different types of defects ranging from Stone-Wales\cite{Ref04}, to
adatoms\cite{Ref13,Ref14,Ref15} and
vacancies\cite{Kra_formation,Ref06,El_Barbary,Ref08,Krasheninnikov_bending,Ref10,Ref11,Ref12,amorim,Krasheninnikov_review,rochaprb}.
Recently, Saito {\it et al.}\cite{saito_multi} have reported on the existence of point defects in
graphene, ranging from a single vacancy to an octovacancy\cite{nota1}. The authors also show that the divacancy, the tetravacancy, and the hexavacancy, are
the most stable defects in a graphene sheet.

In order to highlight the stability of specific vacancies one introduces the concept of magic 
numbers\cite{saito_multi}. These magic numbers indicate the size - the number of atoms removed -
 of a defect  that lead to the most stable multivacancies. Previous studies using positron 
annihilation in graphite suggest that the hexavacancy, $V_{6}$, was the most stable defect\cite{Ref10}. 
This stability was explained by the dangling bond counting model (DBCM), where the number of dangling 
bonds (DB) of the system, $N_{DB}$, decreased as the vacancy became more stable. This model 
was also successful in explaining the stability of multivacancies in silicon\cite{Ref18} and GaAs\cite{Ref19}.

In graphene, however, Saito {\it et al.}\cite{saito_multi}, noted that the stability of the 
defect also depends on pentagons formed upon reconstruction of the system. Subsequently extending 
the dangling bond counting model - adding to that the effect of the pentagons - the authors 
proposed the pentagon and dangling bond counting model (PDBCM). The PDBCM is based on the average 
energy per DB and per pentagon in all defects considered. The model was then used in explaining the 
stability  of vacancies (and consequently the existence of magic numbers) in graphene.

The aim of the present study is twofold. The first one is to determine what are the most stable vacancies in
carbon nanotubes and whether the PDBCM is applicable in the presence of hybridization between $\pi$ orbitals due to
curvature effects.

Secondly and most importantly we propose a modified model based on the PDBCM that uses only two defects 
for the construction of the model. In other words, one does not need to perform calculations for 
all vacancies to determine the parameters for the model; all the parameters - as we will show - 
can be obtained by two preliminary calculations. Most importantly, as it will be demonstrated, 
it contains all the physical ingredients to predict the existence of magic numbers 
in carbon nanotubes, even in the case of large curvature effects.

\section{Method}\label{metod}

{\it Ab initio} total energy calculations based on density functional theory\cite{HK,KS} were 
performed for different types of vacancies on $(5,5)$, $(7,7)$, $(9,9)$ and $(10,10)$ carbon nanotubes. 
This smaller-radius nanotubes was chosen in order to give rise to large curvature effects, 
as oposed to the planar graphene sheet.We used the generalized gradient approximation (GGA)\cite{Perdew_Wang} 
for the exchange and correlation potential within the Perdew-Burke-Ernzerhof
approach \cite{PBE}. Our simulations were performed using a plane-wave DFT method within the VASP
code\cite{kresse1,kresse2} and with ultrasoft pseudopotentials.\cite{vanderbilt} In all our
calculations a plane wave energy cutoff of  $290~eV$ and $8$ k-points along the reciprocal axis of the CNT were used.

\begin{figure}[!h]
\center
  \includegraphics[width=8.5cm]{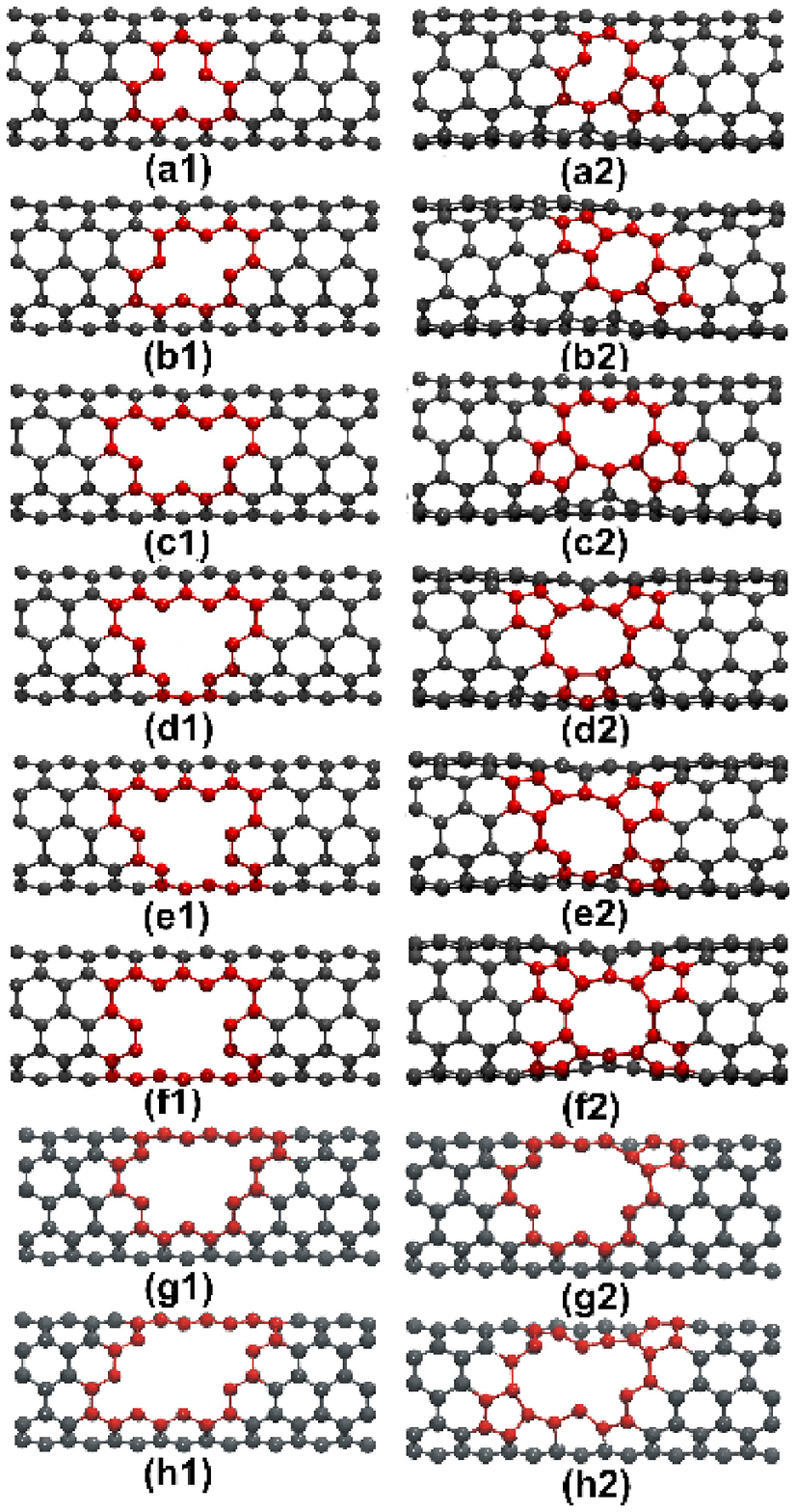}\\
  \caption{\small{Unrelaxed (left hand-side panel) and relaxed (right hand-side panel) multivacancies:
  (a) single vacancy, $V_{1}$; (b) divacancy $V_{2}$; (c) trivacancy $V_{3}$
  (d) tetravacancy $V_{4}$; (e) pentavacacy, $V_{5}$; (f) hexavacancy, $V_{6}$; (g) heptavacancy, $V_{7}$;
  (h) octavacancy, $V_{8}$.}}\label{multirelax}
\end{figure}

Initially the defects are created by simply removing $n$ ($n=1\dots 8$) carbon atoms from a pristine structure containing $8$
irreducible unit cells for each nanotube\cite{nota2}. The $(5,5)$ nanotubes with different numbers of carbon atoms removed prior to relaxation are shown in figure \ref{multirelax}$(a1-h1)$ - left hand-side panel (
the defects on the other nanotubes are equal).
Hereafter we label $V_n$ the respective vacancy where $n$ carbon atoms have been removed.  We note that
the creation of vacancies leads to the formation of dangling bonds - carbon atoms with two-fold coordination
instead of the expected three-fold one - which are energetically unfavorable.
The systems are allowed to atomically rearrange using a conjugate gradient method (CG) until the
forces on all the atoms are lower than $0.02~eV/$\AA. The final relaxed structures are shown in
figure \ref{multirelax}$(a2-h2)$ - right hand-side panel. One can notice that the defects undergo a
reconstruction that leads to the formation of pentagons and subsequent saturation of the dangling bonds. 
We note that in the case of $n$ even, the number of DBs goes to zero (except for $V_{8}$).

\section{Results}

The formation energy, $E_{f}[n]$ for the $V_{n}$ vacancy is calculated using,
\begin{eqnarray}\label{formation_energy}
  E_{f}[n]=E_{r}[n]-E_{p}\left[n\right]-n\mu_{C} ~,
\end{eqnarray}
where $E_{p}$ is the total energy of the pristine nanotube, $E_{r}[n]$ is the 
total energy of the reconstructed nanotube with a $V_{n}$ defect, $n$ is the number of 
carbon atoms removed from the system and $\mu_{C}$ is the chemical potential of $1$ carbon atom, 
which is the total energy of the pristine nanotube divided by the number of atoms on the sysmtem.

\begin{figure}[!h]
\center
  \includegraphics[width=8.5cm]{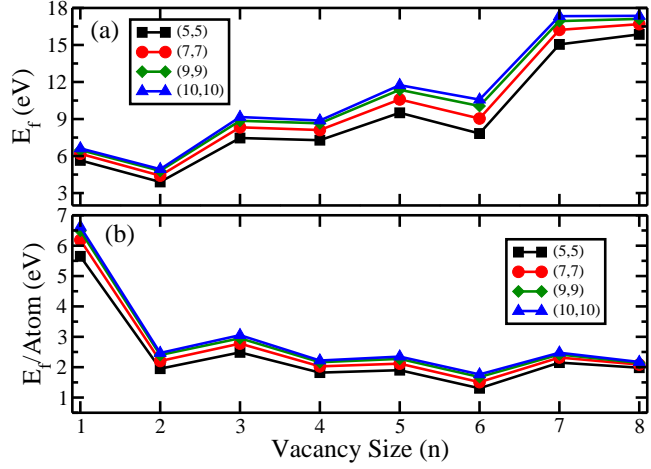}\\
  \caption{\small{(a) Total formation energy, and (b) formation energy per carbon atom removed for a $(5,5)$,
  $(7,7)$, $(9,9)$ and $(10,10)$ carbon nanotubes obtained via DFT calculations.}}\label{form_energy_defects}
\end{figure}

The total formation energy as a function of vacancy size is shown in figure \ref{form_energy_defects}$a$.
From the figure, one might be tempted to say that the $V_{2}$ is the most stable vacancy. 
However, a more reasonable approach is to compare each vacancy normalizing the number of atoms that 
have been removed from the system\cite{nota3}. In that case we observe from figure \ref{form_energy_defects}$b$ that the hexavacancy has the lowest formation 
energy per $C$ atom removed. We also note that $E_f\left[n\right]$ is nonmonotonic. Instead, 
it has local minima for even $n$. This result is in line with previous results for graphene 
indicating the existence of the so-called magic numbers, namely $2$, $4$ and $6$ for carbon nanotubes as well as graphene.

That, however, is not the full picture. One step further into fully understanding the stability of 
vacancies in CNTs is to determine how they are correlated with closely sized vacancies, 
for instance, whether a $V_{6}$ will break into a single  vancancy, $V_{1}$, and a pentavacany, $V_{5}$, and so on.

Hence, in order to address defect stability Saito {\it et al.}\cite{saito_multi} proposed two quantities 
associated with distinct dissociation processes. The first one assumes that a $V_{n}-type$ 
defect breaks up into a single vacancy, $V_{1}$, and a $V_{n-1}-type$ vacancy,

\begin{eqnarray}\label{reac_diss01}
  V_{n}\rightarrow V_{n-1}+V_{1} ~.
\end{eqnarray}
In that manner the first dissociation energy is defined as the energy change between initial and final states,

\begin{eqnarray}\label{eqdiss_en01}
  D_{1}\left[n\right]=E_{f}\left[n-1\right]+E_{f}\left[1\right]-E_{f}\left[n\right] ~.
\end{eqnarray}

For the second case, one would have two $V_{n}-type$ defects that reconstruct into a $V_{n-1}-type$
and a $V_{n+1}-type$ vacancy. In other words, a single vacancy breaks away from one of the defects and
migrates to another one close by

\begin{eqnarray}
  2V_{n}\rightarrow V_{n-1}+V_{n+1} ~.
\end{eqnarray}

Thus the second dissociation energy considered here is defined as

\begin{eqnarray}\label{eqdiss_en02}
  D_{2}\left[n\right]=E_{f}\left[n+1\right]+E_{f}\left[n-1\right]-2E_{f}\left[n\right] ~.
\end{eqnarray}

We finish by highlighting that, following the above definition, the higher the value of $D_1\left[n\right]$
and $D_2\left[n\right]$, the more stable the defect is.

\begin{figure}[!h]
\center
  \includegraphics[width=8.5cm]{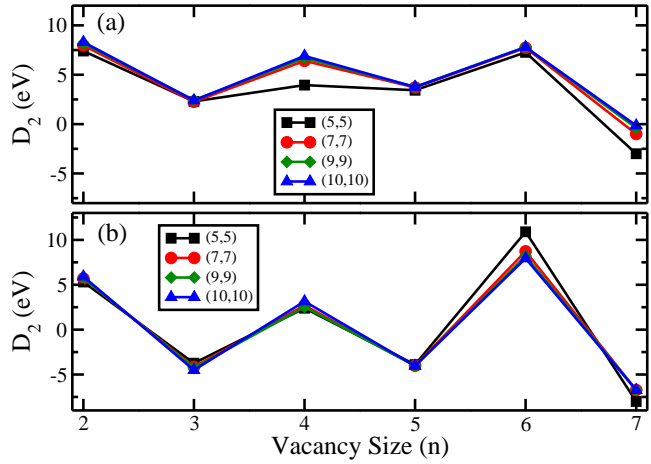}\\
  \caption{\small{Dissociation energies (a) $D_1\left[n\right]$ (equation (\ref{eqdiss_en01})), and (b)
  $D_2\left[n\right]$ (equation (\ref{eqdiss_en02}))} as a function of defect size for a $(5,5)$, $(7,7)$, 
  $(9,9)$ and $(10,10)$ carbon nanotubes.}\label{diss_ener}
\end{figure}

In Figure \ref{diss_ener}$(a-b)$ we observe peaks in the dissociation energy, localized in
$V_{2}$, $V_{4}$ and $V_{6}$. Following the same idea given by Saito {\it et al.}\cite{saito_multi}, we
can conclude that the vacancies $V_{2}$, $V_{4}$ and $V_{6}$ are stable in the carbon nanotubes
in a fashion similar to graphene, corroborating the initial conclusion that there are three magic numbers. 
Furthermore, the curvature for such nanotubes is quite large, therefore we can also conclude that 
these magic numbers are valid for graphene and nanotubes in general.

In order to understand the origin of the stability for these particular defects  we will focus on the  
nanotube that has the largest curvature effect - the $(5,5)$. First we will look to a model that only takes into account the dangling bonds, 
namely the dangling bond counting model $(DBCM)$\cite{saito_multi}. We therefore need to determine some quantities. 
The first is the gain in energy due to the relaxation of the system, which is defined as the difference 
in energy between the reconstructed system and  the energy of the system with the unrelaxed defect, $E_u$,

\begin{eqnarray}
E_{relax}\left[n\right]=E_u\left[n\right]-E_r\left[n\right]~.
\end{eqnarray}

The pentagon bond energy is then defined as the ratio between the relaxation energy and the number of pentagons,
 $N_{Pent}$, in the relaxed CNT

\begin{eqnarray}
E_{pent}\left[n\right] = E_{relax}\left[n\right]/N_{pent} ~.
\end{eqnarray}

Finally we also define the energy per dangling bond as the ratio between the formation
energy of the unrelaxed defect and the number of dangling bonds

\begin{eqnarray}
E_{DB}\left[n\right]=\left(E_u-E_p\left[n\right] - n \mu_C \right)/N_{DB} ~.
\end{eqnarray}

In Table \ref{tab_ener} we summarize these quantities for each one of the vacancies in the present study obtained from our DFT calculations.

\begin{table}[!h]
 \vspace{3mm} \centering
  \begin{tabular}{cccccccccc}
  \hline\hline
     & $V_{1}$ & $V_{2}$ & $V_{3}$ & $V_{4}$ & $V_{5}$ & $V_{6}$ & $V_{7}$ & $V_{8}$ & $\langle V\rangle$ \\
    \hline \\
   $E_{relax}$ $(eV)$ & $2.305$ & $5.51$ & $5.04$ &  $7.06$ & $7.18$ & $10.13$ & $2.61$ & $5.04$ & - \\
    $E_{pent}$ $(eV)$ & $2.30$ &  $2.75$ & $2.52$ & $2.35$ & $2.39$ & $2.53$ & $2.61$ & $2.52$ & $2.50$\\
    $E_{DB}$ $(eV)$ & $2.64$ & $2.35$ & $2.50$ & $2.39$ & $2.38$ & $2.25$ & $2.74$ & $2.67$ & $2.49$ \\
    $N_{pent}$ $(eV)$ & $1$ & $2$ & $2$ & $3$ &  $3$ & $4$ & $1$ & $2$  & - \\
    $N_{DB}$ & $3$ & $4$ & $5$ & $6$ & $7$ & $8$ & $7$ & $8$  & - \\
    $N_{DB}^{relax}$ & $1$ & $0$ & $1$ & $0$ & $1$ & $0$ & $2$ & $0$  & - \\
    $N_{Sides}$      & $9$ & $8$ & $10$ & $9$ & $11$ & $10$ & $17$ & $16$ & - \\
  \hline\hline
    \end{tabular}
    \caption{\small{Energies associated with atomic relaxations, $E_{relax}$, pentagons, $E_{pent}$, dangling bonds, $E_{DB}$ for each type of vacancy. The total number of pentagons, $N_pent$, dangling bonds (prior to, $N_DB$ and after, $N_{DB}^{relax}$, reconstruction), and the number of sides, $N_sides$ of the reconstructed poligon  are also shown. The last column indicates the average of some of these quantities over all the defects.}}
    \label{tab_ener}
\end{table}

The first model we analyze only takes into consideration the dangling bonds in the system. 
Thus the formation energy in the so called dangling bond counting model is simply given by
\begin{equation}\label{efor_DBCM}
  E_f^{DBCM}\left[n\right]=2.49N_{DB}
\end{equation}
where $N_{DB}$ is the number of dangling bonds for the unreconstructed system and the 
proportionality factor corresponds to the energy per DB, averaged over all possible vacancies as shown in Table \ref{tab_ener}.

The DBCM is a rather simplified model and it does not take into account the strong relaxation of the defects.
On the other hand the pentagon and dangling bond counting model proposed by Saito {\it et al.} also considers
the reconstruction of the vacancies into pentagons. In the PDBCM the formation energy is 
thus obtained by considering also the reconstruction of the vacancies into pentagons. 
Hence the pentagon and dangling bond counting model adds a correction to the DBCM to include the gain in 
energy due to the formation of pentagons. This new term includes the number of pentagons 
and the average pentagon bond energy. Thus the formation energy in the PDBCM is written as
\begin{equation}\label{eq_pdbcm}
  E_f^{PDBCM}\left[n\right]=2.49N_{DB}-2.50N_{Pent}
\end{equation}
where $ N_{PB}$ is the number of pentagons of the relaxed system. We note that the formation 
energy is now given by a competition between the high-energy dangling bonds and the 
reconstruction of the defects into pentagons to try to eliminate as many  DBs as possible.

\begin{figure}[!h]
\center
  \includegraphics[width=8.5cm]{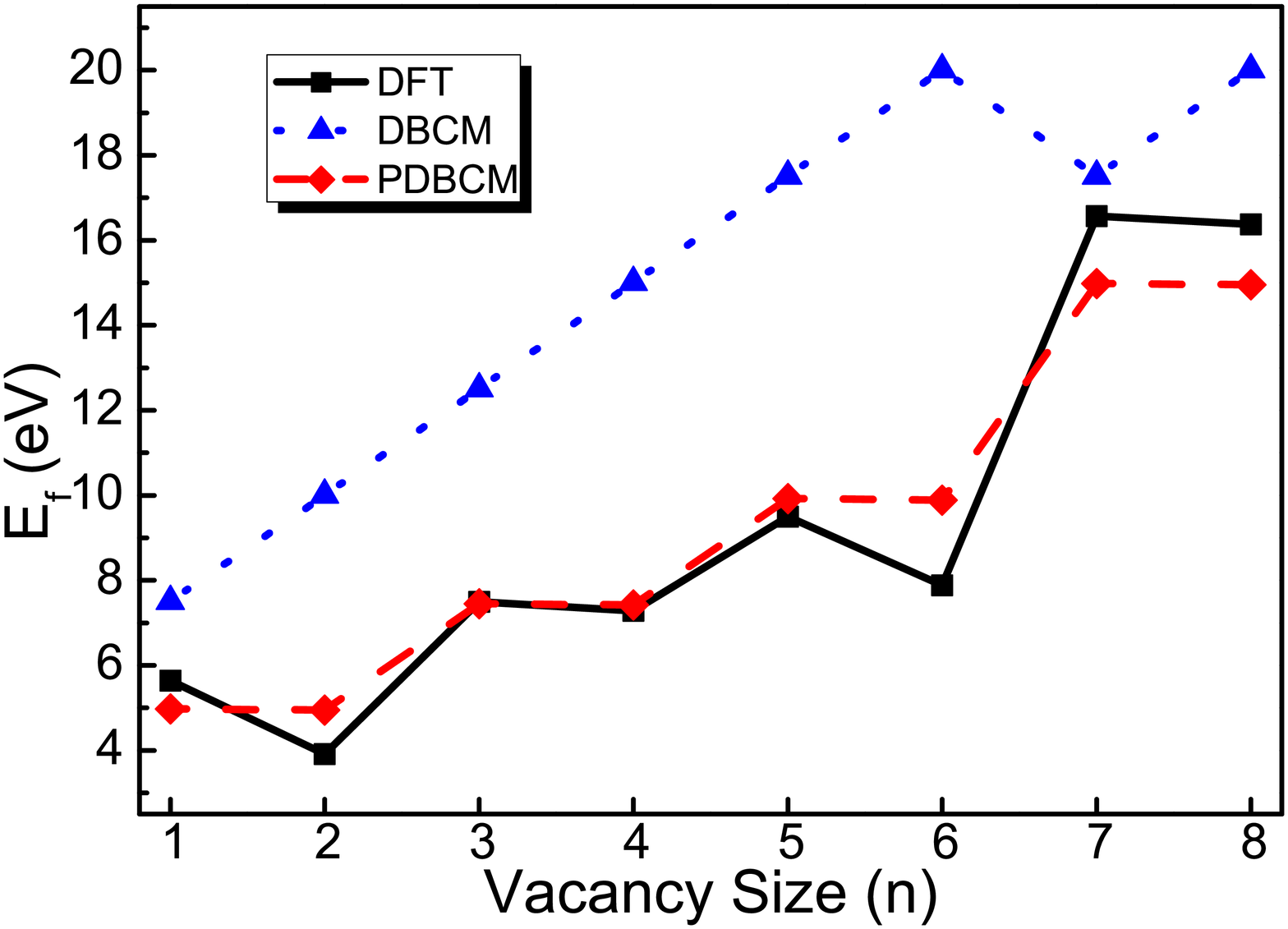}\\
  \caption{Formation energies calculated using the DBCM (blue triangle) and PDBCM (red diamonds). The
  DFT calculations using equation \ref{formation_energy} (black square) are used as a benchmark for the
  quality of both models}\label{form_all_models}
\end{figure}

In figure \ref{form_all_models} we present the formation energy calculated with both methods, 
DBCM and PDBCM, compared with our DFT results assumed here to be a benchmark calculation.

From  figure \ref{form_all_models} we can extract two pieces of information. First, if we consider 
only the dangling bonds, we obtain a linear relationship over a wide range of defects. 
That is clearly not the profile seen by performing the full DFT calculation. It is then 
clear that the reconstruction plays an important role in the formation energy of the defect. 
Using the PDBCM, where not only the dangling bonds are taken into consideration but also the 
number of pentagons formed after relaxation, we can see that the results are in 
reasonable agreement with our DFT calculations.

\begin{figure}[!h]
\center
  \includegraphics[width=8.5cm]{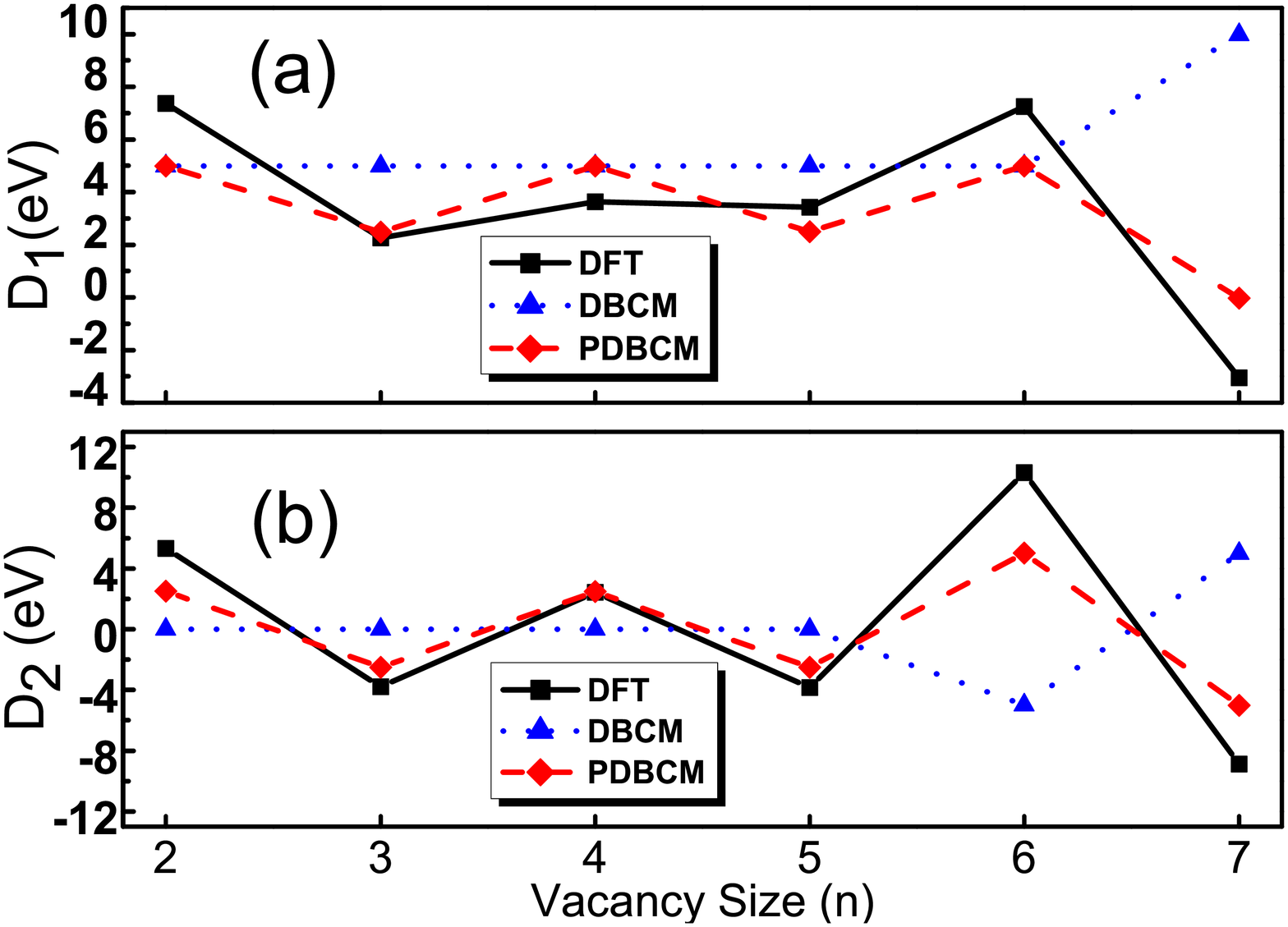}\\
  \caption{Dissociation energies a) $D_1\left[n\right]$ (equation (\ref{eqdiss_en01})), and b) $D_2\left[n\right]$ (equation (\ref{eqdiss_en02})) for the  dangling bond couting model and the pentagon and dangling bond counting model. We also present our DFT results for comparsion.}\label{diss_ener_models}
\end{figure}

We also use both models to fit the dissociation energies and to determine whether they 
reproduce the magic numbers that have been found in our DFT calculations. From figure \ref{diss_ener_models} - 
which shows the dissociation energies as a function of defect size - we can conclude again, 
that the dangling bond counting model fails to reproduce the two dissociation energies, whereas the 
model that takes into account the pentagons, reproduces reasonably well (specially in the case of $D_{2}$) our DFT calculations.

Thus, the PDBCM as proposed by Saito {\it et al.}, can be used to explain the appearance of the 
magic numbers and the stability of the vacancies. The model presented here uses averages 
over all defects, but even if one uses only two defects - the single vacancy and the $V_{2}$ - 
would lead to a similar result. One downside of the PDBCM, however, is the fact that it cannot 
account for the dips in the formation energy for even-numbered vacancies. By construction, the 
formation energies of these vacancies are always identical to the previous odd-numbered ones. 
Thus, $E_f\left[n\right]$, it is not possible to infer the existence of the magic numbers. 
This leads to the conclusion that the model is missing an important ingredient. In the light 
of this problem we propose a new model for the formation energy of multivacancies. This model 
retains the spirit of the PDBCM where DBs and the reconstruction of the defect are the main 
ingredients for the formation energy, but it should also include information about the final 
shape of the defect. In others words, energy is gained by forming the pentagons, but in 
detriment of other bonds that are bent after the relaxation is complete.

Our proposal for the formation energy is based on three main contributions. The first includes 
the contribution due to the dangling bonds while the second accounts for the formation of  
pentagons after relaxation. Both of which retain the same spirit of the PDBCM,
\begin{equation}\label{eq_ourmodel}
  E_{f}[n]=E_{DB}N_{DB}-E_{P}N_{P}+E_{S}(N_{S}-N_{P}).
\end{equation}
The third and final term accounts for the relaxation energy $(E_{s})$ of all bonds in the 
defect which have not been included in the pentagons.

In order to determine the parameters described above, we choose the single vacancy and the trivacancy
\cite{nota4}. Since both show, upon relaxation, pantagons and dangling bonds. The dangling bond energy is 
obtained by taking the average between two defects. We then use the DFT results for $V_{1}$ and $V_{3}$ in 
equation (\ref{eq_ourmodel}) together with the average dangling bond energy. This leads to a 
system of simultaneous equations if one assumes the DFT result to be the correct formation 
energy in this case. The parameters are thus summarized in table \ref{tab_enermodel}.

\begin{table}[!h]
 \vspace{3mm} \centering
  \begin{tabular}{ccc}
  \hline\hline
    $E_{DB}(eV)$ & $E_{P} (eV)$ & $E_{S}(eV)$ \\
    \hline\hline
   $2.57$ & $-3.28$ & $0.15$ \\
  \hline\hline
    \end{tabular}
    \caption{\small{Parameters used in our calculations of the formation energies of $n-$vacancies in a $(5,5)$  carbon nanotube.}}
    \label{tab_enermodel}
\end{table}

\begin{figure}[!h]
\center
  \includegraphics[width=8.5cm]{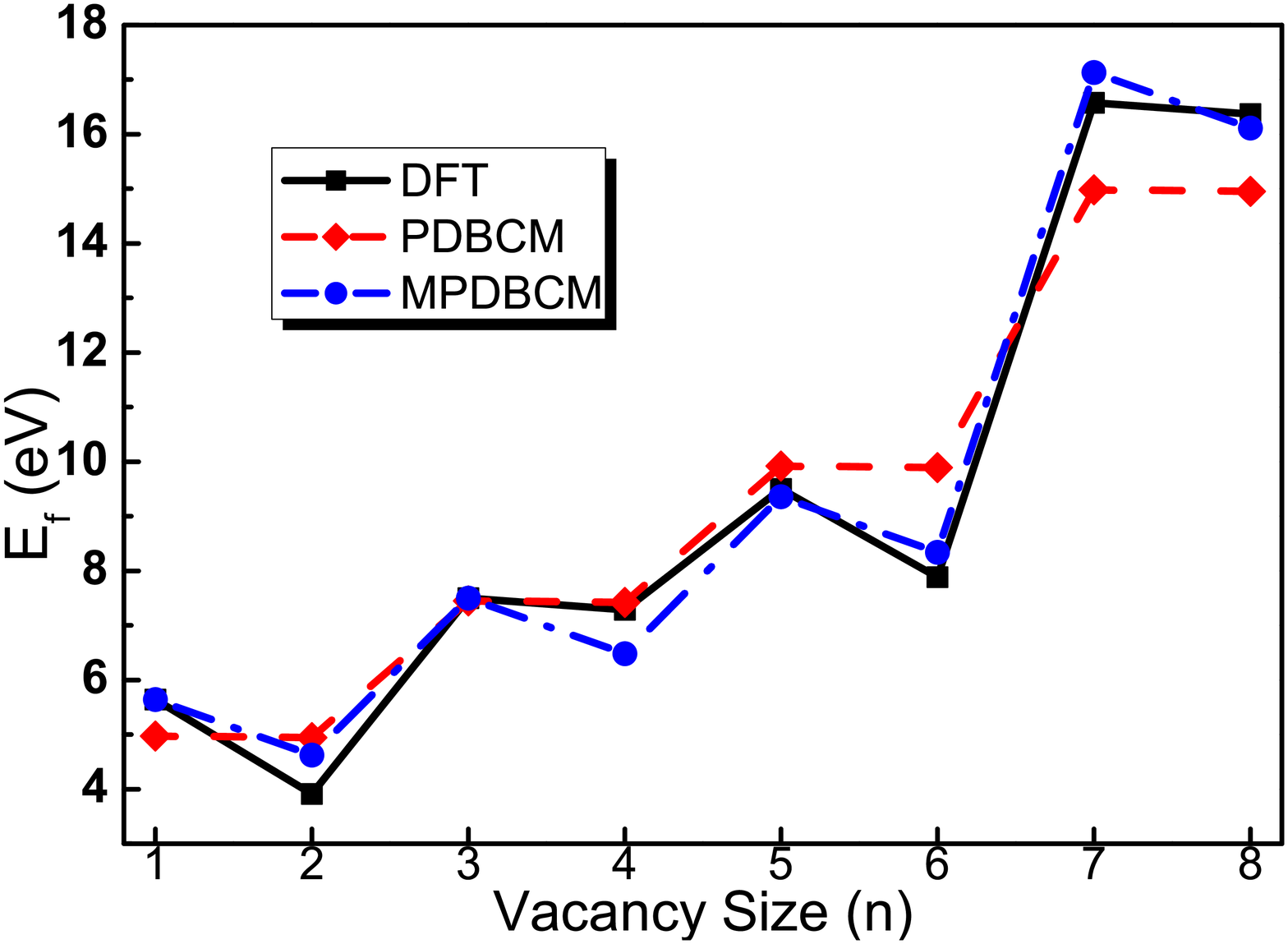}\\
  \caption{\small{Formation energies calculated with the Modified-PDBCM (equation (\ref{eq_ourmodel})),
  and with the PDBCM  (equation (\ref{eq_pdbcm})). Our benchmark DFT calculations (black square)
  are also shown for the sake of comparison.}}\label{former_modelos}
\end{figure}

The formation energies for all defects, calculated from equation (\ref{eq_ourmodel}) are depicted
in figure \ref{former_modelos}. One can see that the MPDBCM is, in general, better than the PDBCM. In particular
it is capable of accounting for the dips on the formation energy of even-numbered multivacancies.
The only case where the MPDBCM performs worse is the tetravacancy. For the $V_{4}$, the pentagon oriented along
the tube has much longer bond lenghts when compared to the $C-C$ bonds ( $1.62\AA$ as oposed to $1.50\AA$). 
This leads to a discrepancy in the formation energy that coincidentally is correct in the PDBCM.

\begin{figure}[!ht]
\center
  \includegraphics[width=8.5cm]{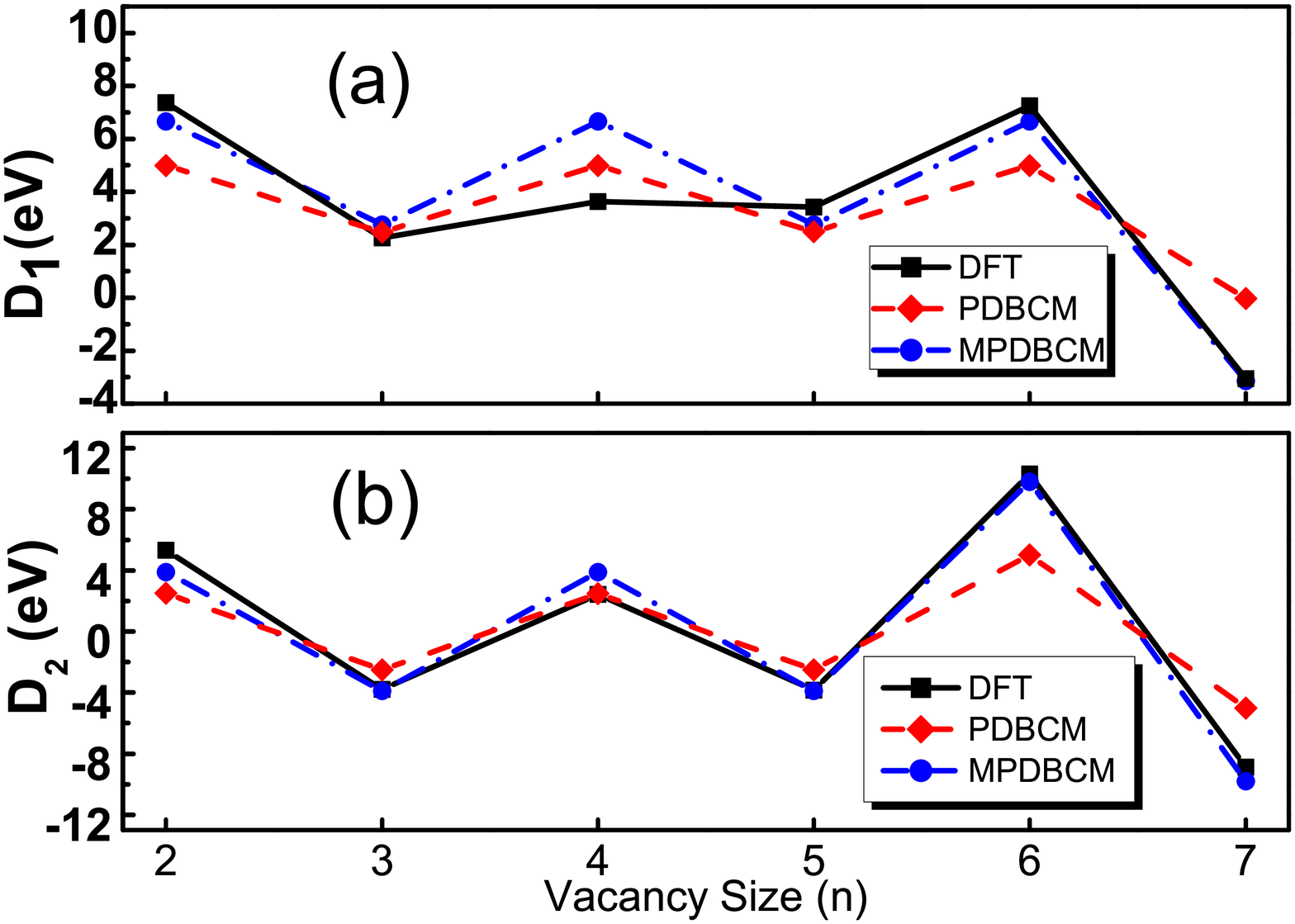}\\
  \caption{Dissociation energies as a function of the number of vacancies. a) $D_1\left[n\right]$ and b) $D_2\left[n\right]$
   }\label{diss_en_ze}
\end{figure}

The dissociation energies calculated with the MPDBCM (shown in figure \ref{diss_en_ze}) are also in very
good agreement when compared with our DFT calculations. Again they are usually better.

\section{Conclusions}

Hence we have calculated the formation and dissociation energies associated with $n$-vacancies in carbon nanotubes. We have demostrated that in carbon nanotubes with large curvature effects as in graphene the same stable multivacancies appear, namely the divacancy, the tetravacancy and the hexavacancy. The existance of these magic numbers is corroborated by the pentagon and dangling bond counting model which shows that the stability of the defects is given by the competition between high energy dangling bonds and the reconstruction of the multivacancies. In the cases where the defects are most stable we note that the relaxation leads to no dangling bonds left.

Finally we proposed a modified method based on the PDBCM which also takes into account the final shape of the system. It takes into consideration both dangling bonds and the reconstruction of the defect and adds to that effect of bond streching and bond bending that takes place of reconstruction. This way we are able to include one important physical ingredient that was missing from previous models. From our calculations one can obtain the parameters that fit extremely well the formation and dissociation energies of a number of $n$-vacancies. In particular, our model is capable of accounting for magic numbers 2,4 and 6 which are related to the most stable defects.

\bibliographystyle{apsrmp}

\begin{thebibliography}{35}
\bibitem{nota2} In CNTs, as opposed to graphene, there is sometimes more than one non-equivalent
way of making an $n$-vacancy due to curvature effects. In this work we only consider, for each $n$ vacancy, the
arrangement with the lowest formation energy.
\bibitem{nota3} In a sense, that would be equivalent to answering
whether it is easier to create $n$ single vacancies or only one $n$-vacancy.
\bibitem{nota4} The reason for this choice is that both defects have pentagons, dangling bonds and a
final form that is similar, with a difference that the trivacancy have one pentagon more than the single vacancy.
\end{thebibliography}

\end{document}